\documentclass[preprint,aps,graphicx]{revtex}

\def\qq{{\kern0.05em\hbox{\textsf{q}}}}

\def\II{{\cal I}(u \leftrightarrow d)}

\def\red{
}
\def\black{
}

\def\mycomm#1{\hfill\break\strut\kern-3em{\red\tt ====> #1\black}\hfill\break}

\def\eqref#1{(\ref{#1})}

\makeatletter
\def\hlinewd#1{\noalign{\ifnum0=`}\fi
\hrule \@height #1 \futurelet \reserved@a\@xhline}
\def\hwhiteline{\noalign
{\ifnum0=`}\fi\hrule
\@height 0pt\vskip 1.0ex\futurelet \reserved@a\@xhline}
\makeatother

\def\black{\special{ps: 0.0 setgray}}

\newcommand{\mydraft}{
\newcount\timecount
\newcount\hours \newcount\minutes  \newcount\temp \newcount\pmhours

\hours = \time
\divide\hours by 60
\temp = \hours
\multiply\temp by 60
\minutes = \time
\advance\minutes by -\temp
\def\hour{\the\hours}
\def\minute{\ifnum\minutes<10 0\the\minutes
    \else\the\minutes\fi}
\def\clock{
\ifnum\hours=0 12:\minute\ AM
\else\ifnum\hours<12 \hour:\minute\ AM
\else\ifnum\hours=12 12:\minute\ PM
    \else\ifnum\hours>12
     \pmhours=\hours
     \advance\pmhours by -12
     \the\pmhours:\minute\ PM
     \fi
    \fi
\fi
\fi
}
\def\fullclock{\hour:\minute}
\begin{centering}
\red
\font\Hugett  =cmtt12 scaled\magstep4
\hbox{\Hugett Draft:\today,\clock}
\black
\end{centering}
\vskip -1.7cm
$\phantom{a}$
} 

\def\beq#1{\begin{equation} \label{#1}}
\def\eeq{\end{equation}}
\def\bra#1{\langle #1\vert}
\def\ket#1{\vert #1\rangle}

\def\bea{\begin{eqnarray}}
\def\eea{\end{eqnarray}}

\newskip\humongous \humongous=0pt plus 1000pt minus 1000pt

\newif\ifdtup

\begin{document}

{\tighten

\title{\boldmath
Isospin Violation in $X(3872)$:
\\
Explanation From a New Tetraquark Model
\unboldmath}

\author{Marek Karliner\,$^{a}$\thanks{e-mail: \tt marek@proton.tau.ac.il}
\\
and
\\
Harry J. Lipkin\,$^{a,b}$\thanks{e-mail: \tt
ftlipkin@weizmann.ac.il}
}

\address{ \vbox{\vskip 0.truecm}
$^a\;$School of Physics and Astronomy \\
Raymond and Beverly Sackler Faculty of Exact Sciences \\
Tel Aviv University, Tel Aviv, Israel\\
\vbox{\vskip 0.0truecm}
$^b\;$Department of Particle Physics \\
Weizmann Institute of Science, Rehovot 76100, Israel \\
and\\
Physics Division, Argonne National Laboratory \\
Argonne, IL 60439-4815, USA\\
}
\maketitle
\begin{abstract}
New data for $X(3872)$ production in $B$ decays provide a separation
between $X$ production and decay, sharpen several experimental puzzles
and impose serious constraints on all models. Both charged and
neutral $B$ decays produce a narrow neutral resonant state that decays to
both $J/\psi \rho$ and $J/\psi \omega$, while no charged resonances
in the same multiplet are found. This suggests that the $X$ is an
isoscalar resonance whose production conserves isospin, while isospin
is violated only in the decay by an electromagnetic interaction allowing
the isospin-forbidden $J/\psi \rho$ decay. A tetraquark isoscalar $X$
model is proposed which agrees with all present data, conserves isospin in
its production  and breaks isospin only in an electromagnetic $X(3872)
\rightarrow J/\psi\,\rho^o$ decay. The narrow $X$ decay width results
from the tiny phase space available for the $J/\psi \omega$ decay and
enables competition with the electromagnetic isospin-forbidden $J/\psi
\rho $ decay which has much larger phase space. Experimental tests
are proposed for this isospin production invariance.
\end{abstract}
\pagebreak
\section {The Puzzle - How is isospin broken in X(3872)?}
\subsection {Is isospin broken in production of X(3872) or in its decay?}

The decays of the baffling resonance  $X(3872)$
\cite{Choi:2003ue,Acosta:2003zx,Abazov:2004kp,Aubert:2004ns,PDG,Godfrey:2009qe}
have raised puzzles leading to
suggestion that it breaks isospin symmetry\cite{nils,oset}.
In particular, both $X(3872)\rightarrow J/\psi \pi\pi$ and
$X(3872)\rightarrow J/\psi \omega$ decay modes have been observed
\cite{Abe:2005ix,delAmoSanchez:2010jr}.
Since isospin
symmetry has been shown to be an excellent symmetry of strong interactions and
$SU(3)_{flavor}$ been shown to be a good approximate symmetry, we investigate
ways to solve this puzzle.

New data on $B\rightarrow KX$ sharpen this puzzle.
The weak $B$ decay provides an unambiguous production
mechanism.
The dominant $b$  quark decay into charmonium states goes via the vertex

\beq{vertex}
b \rightarrow c \bar c s
\eeq
Although the weak interaction generally violates isospin, this vertex
conserves isospin. The $b$, $c$ and $s$ flavors are all isospin scalars.
The decay of a $B$ meson state $B(\bar b\qq)$ in which a $\bar b$ antiquark
is bound to a nonstrange quark of flavor $\qq$ is a $\delta I =0$ transition.
If the strong interaction following the weak decay is invariant under
the interchange of the $u$ and $d$ flavors,
denoted by $\II$, we have
\ \ $\II \ket{B^+} = \ket{B^o}$ \ and
we find the transmission matrix elements $\bra{f}T\ket{B}$ for any B decay
into any final state denoted by $\ket{f} $ satisfy the relation
\beq{reflec}
\bra{f}T\ket{B^+}= \bra {\,\II f\,}T\ket{B^o}
\eeq
    The new data violate isospin invariance, also called charge independence,
which is a continuous $SU(2)$ transformation in the $u,d$ space.
But they seem to satisfy the
isospin reflection (\ref{reflec}) which is a discrete subgroup of $SU(2)$ also called
charge symmetry and is less than charge independence. For example, the charge symmetry
reflection relates the
pair of decays $B^+ \rightarrow J/\psi K^+ \rho^o$ and $B^o \rightarrow J/\psi K^o \rho^o$
which can go via a neutral $X$. Charge symmetry also relates the pair
$B^+ \rightarrow J/\psi K^o \rho^+$ and $B^o \rightarrow J/\psi K^+\rho^-$
which cannot go via a neutral $X$.
Full charge independence of isospin invariance relates all four decays and is violated by
the absence of a charged $X$.

This imposes serious constraints on all models for the X which are not
easily satisfied.
None of the proposed models for the $X$ can explain  all the following
three observations

\begin{itemize}
\item[1.]
The $X$ decays into both into the isovector $J/\psi \rho^o$ and
the isoscalar $J/\psi \omega$
\cite{Abe:2005ix,delAmoSanchez:2010jr}.

\item[2.]
No charged $X$ partner decaying into the charged partner
of $J/\psi \rho^o$ has been found \cite{Aubert:2004zr}.

\item[3.]
The $X$ is produced at comparable rate
both in charged and neutral $B^+\rightarrow K^+X_u$
and $B^o\rightarrow K^oX_d$ decays \cite{Aubert:2008gu},
where the states denoted by
$X_u$ and $X_d$ are either the same state or two states satisfying
$X_u =  \II\, X_d$.

\end{itemize}

None of the proposed models can explain how both charged and neutral $B$
mesons can produce a neutral resonant state that decays to both $J/\psi \rho$
and $J/\psi \omega$, while no charged resonances in the same multiplet
are found. The models which use a $D^o\bar D^{*o}$ ``deuson"\cite{nils} contain no $d$
flavored quarks and are not easily produced produced from a neutral $B$
which has a spectator $d$ quark.
That already says that there is a problem which none of the proposed models can solve.

We now see that the new data raise the question of whether the isospin breaking  occurs in the
production of the resonance, in the propagation in time of the resonance or in the decay
process.
In the remainder of this paper we will show that all the problems listed above can be
solved in a model that assumes the $X$ is isoscalar and has no charged partner, that its
production and propagation in time all
satisfy isospin invariance, and all the isospin breaking occurs only in its decays
into the isovector $\,J/\psi \rho^o\,$ which can be electromagnetic.

\subsection {A nonet structure for $c \bar c q \bar q$ states in $b$ decays}

A charmonium $c \bar c$ quark pair is an $SU(3)_{flavor}$ singlet whose strong
interactions with $u$, $d$ and $s$ quarks are flavor independent. We therefore
assume that the  strong dynamics of a light quark system  and its flavor
structure will be negligibly changed by the addition of a charmonium pair. Thus
we assume that the spectrum of a system of a light quark $q \bar q$ pair and a
a $c \bar c$ will  have the nonet structure of the light quark symmetry with
flavor symmetry broken only by the mass difference between the strange and
nonstrange quarks.

$B$ decays via the vertex
(\ref{vertex}) into charmonium and two light quark mesons denoted by $M_1$ and
$M_2$ are therefore described by the diagrams
\beq{KM12}
B (b\bar \qq) \,\rightarrow\, c \bar c s \bar \qq \,\rightarrow\,
(c \bar c) s \bar \qq \, q \bar q\, \rightarrow\,
M_1(s \bar q) \, (c \bar c) \, M_2(q \bar \qq)
\eeq
\beq{KM12b}
B (b\bar \qq) \,\rightarrow\, c \bar c s \bar \qq \,\rightarrow\,
(c \bar c) s \bar \qq \, q \bar q\, \rightarrow\,
M_1(s \bar \qq) \, (c \bar c) \, M_2(q \bar q)
\eeq
where the $\bar \qq$ is the spectator antiquark and $(q \bar q)$ denotes a
flavor singlet quark-antiquark pair created by gluons.
The diagram (\ref{KM12b}) where the $q \bar q$ pair appears in the same
$M_2$ final meson violates the well established OZI selection
rule\cite{Okubo,Zweig,Iizuka} which forbids a quark-antiquark pair created by gluons
to appear in the same final meson.

We are interested in decays of the form
$B \ \rightarrow \ K\, X(3872) \rightarrow \ K\,J/\psi \rho/\omega$.
In these decays
$M_2$ is a light quark neutral vector meson and
the only contributing diagrams (\ref{KM12}) which do not violate
the OZI rule are
\beq{neutves}
B(b\bar \qq) \ \rightarrow \
M_1(s \bar \qq) \, (c \bar c) M_2(\qq \bar \qq)
\ \rightarrow \
M_1(s \bar \qq) \, X_{\qq}
\ \rightarrow \
 M_1(s \bar \qq)\, J/\psi\, V_{\qq}
\eeq
Where $V_{\qq}$ denotes the linear combination of neutral vector meson
states with the quark constituents $\qq \bar \qq$ and $X_{\qq}$ denotes
the linear combination of $X(3872)$ that decays into $J/\psi V_{\qq}$.
\beq{rhomeg}
V_u= \frac{\rho+\omega}{\sqrt 2}; ~ ~ ~ V_d= \frac{\rho-\omega}{\sqrt 2}
\eeq
If two amplitudes related by the isospin reflection are equal,
\beq{neutvesu}
A[\,B(b\bar u) \ \rightarrow \ M_1(s \bar u)\, J/\psi\, V_u\,]
\quad = \quad
A[\,B(b\bar d) \ \rightarrow \ M_1(s \bar d)\, J/\psi\, V_d\,]
\eeq
this then leads to the basic puzzle:
{\em
a neutral final state requires that the only components of the $q\bar q$ pair created by gluons
that contribute to the $X$ production must have the same flavor $\qq$
as the spectator quark.}
For example, consider the decay
\beq{chbjrho}
B^- \ \rightarrow \ K^- X(3872) \ \rightarrow \ K^- \,J/\psi \,\rho^o
\eeq
The $B^-$ meson contains a spectator $\bar u$ quark.
The final state contains three
light quarks, one being the original spectator and an additional
$q \bar q$ pair made by gluons. Since gluons are isoscalar, the additional
$q \bar q$ pair must be isoscalar and can not make a $\rho^o$. Thus one of
the two quarks in the final state $\rho^o$
must be the spectator $\bar u$ quark. Since $\rho^o$ is neutral, the other
quark in $\rho^o$ must be the $u$ quark coming from gluons and the $q \bar q$
pair produced by gluons must therefore be a $u \bar u$.
Similarly, in
\beq{neubjrho}
B^o \ \rightarrow \ K^o X(3872) \ \rightarrow \ K^o \,J/\psi
\,\rho^o,
\eeq
the gluons must only make a $d \bar d$ pair.
This is puzzling, because
components of the $q \bar q$ pair with the opposite  nonstrange flavor
should be produced equally by gluons, but these would produce charged states,
for example
\beq{jrhoch}
B^-(b \bar u) \ \rightarrow \ \bar K^o \,J/\psi \,\rho^-(d \bar u)\,.
\eeq
But no such charged narrow states have been seen at the $X$ mass.

However, before  jumping hastily to conclusions about isospin violation,
we note some remarkable kinematics at the $X$ mass. The $X$ mass is very close
to the $J/\psi \omega$ threshold and may even be slightly below. If the $X$ has
isospin zero and a spin and parity that forbid its decay into two pseudoscalar
mesons, the lowest final state allowed by isospin-conserving strong interactions
is $J/\psi \omega$, which has very small phase space. This can
account for the observed narrow width of the isoscalar $X$.
In some models the $X$ can have an isovector partner.
The large
width of the $\rho$ and its large lower mass tail allow the isovector partner of the $X$ to be well above the
$J/\psi \rho$ threshold. The transitions
(\ref{chbjrho}), (\ref{neubjrho}) and (\ref{jrhoch})
will all have a very large phase space and
be very wide for the isovector partner and unobservable against a continuum background.
Only the transition
\beq{neubjomeg}
B^o \ \rightarrow \ K^o X(3872) \ \rightarrow \ K^o \,J/\psi
\,\omega
\eeq
survives with a narrow width for strong decay. However,
even for a purely isoscalar $X(3872)$
 the transitions
(\ref{chbjrho}) and (\ref{neubjrho}) can go via an electromagnetic
isospin breaking interaction. Although electromagnetic decays
are generally much weaker than strong decays, the large
phase space enables these electromagnetic decays to compete with the allowed
strong decay (\ref{neubjomeg}).

This enormous phase space difference must play a role in any model for the X(3872). Models
like the deuson model which have a mixed isospin state for the X(3872) cannot easily explain
why the large phase space for the $J/\psi \rho$ decay does not overwhelm the $J/\psi \omega$
decay. If the transition matrix elements are comparable for the two
decays, the phase space
difference will greatly enhance the $J/\psi \rho$ decay branching ratio above the small $J/\psi \omega$
branching ratio.

Outside the $X$ mass region where phase space is not so different
we can find relations that can be tested by experiment.
OZI-violating transitions involve an isoscalar pair created by gluons which can only produce
the isoscalar $\omega$. We first neglect the OZI violating transition.
In charged $B$ decays
the light neutral vector meson in the final state
appears with the constituents $u\bar u$ which is not an isospin eigenstate,
and in neutral $B$ decays it appears with constituents
$d \bar d$ which also is not an isospin eigenstate.
Thus all neutral nonstrange vector mesons produced in these
decays must be equal
linear combinations of two isospin eigenstates with isospin zero and one, like
$\rho^o$ and $\omega$.

We can include a small correction to this $\rho^o-\omega$ equality resulting from possible OZI violations which add an extra $\omega$
amplitude, but will not affect the $\rho$ amplitude.

For the case where $M_1$ is a pseudoscalar meson eq. (\ref{neutves}) gives
the relation
\beq{als}
A[B^{\pm} \rightarrow J/\psi K^{\pm} \omega] =
A[B^{\pm} \rightarrow J/\psi K^{\pm} \rho^o]
\eeq
This is the analog of the first quark-model OZI relation \cite{alex},
\beq{als2}
A[ K^-p \rightarrow \Lambda \omega] = A[ K^-p \rightarrow \Lambda \rho^o]
\eeq

The relation (\ref{als}) follows from $U(3)$ flavor symmetry
and holds for all regions
of phase space where $U(3)$ symmetry is not broken.

Since isospin is a subgroup of $U(3)$, the  isospin reflection
of eq.~(\ref{als})\, is equally valid and related to eq.~(\ref{als}),
\beq{als3}
A[B^{\pm} \rightarrow J/\psi K^{\pm} \omega] =A[B^o \rightarrow J/\psi K^o \omega] =
A[B^o \rightarrow J/\psi K^o \rho^o] =
A[B^{\pm} \rightarrow J/\psi K^{\pm} \rho^o]
\eeq
In the $U(3)$ symmetry limit all nine $q_i\bar q_j$ mesons are degenerate for all three values
of the flavors $i$ and $j$: \ $i,j=u,d,s$. There are two commonly noted symmetry breaking
mechanisms:
(1) the quark mass differences;
(2) the annihilation diagram where one flavor is
annihilated into gluons and a pair of another flavor is created.

When the annihilation diagram
is dominant, $U(3)$ is broken into $SU(3) \otimes U(1)$, as in the pseudoscalar nonet, where the
$SU(3)$ singlet $\eta'$ is separated from the pseudoscalar octet. The $U(3)$ symmetry
breaking in the pseudoscalars is generally attributed to the anomaly which affects
only the singlet.

     In the vector mesons, the quark mass difference between strange
and nonstrange quarks is dominant and $U(3)$ is broken into $U(2)_{ud}\otimes U(1)_s$.
At this stage the $\rho$ and $\omega$ are degenerate. A small annihilation diagram then
splits the $U(2)$ quartet into an $SU(2)$ triplet $\rho$ and an $SU(2)$ singlet $\omega$.
In electromagnetic interactions where the electric charge difference between the $u$ and
$d$ quarks is more important than the isospin conserving strong interaction, isospin is
broken and the $u \bar u$ and $d \bar d$ states are closer to experiment than the isospin
eigenstates.

However, the equality (\ref{neutvesu})\, indicates a different
mechanism from the $ud$ mass difference for isospin breaking.  That two
quark-antiquark pairs with different nonstrange flavors should be produced
very unequally from gluons would indicate isospin breaking. But this type
of isospin breaking is not predicted by any of the proposed models for
the $X$. The data indicate that ``charge symmetry" $\II$ is still valid
but  ``charge independence" is violated.  The full isospin symmetry
which involves continuous transformations in the $ud$ flavor space is
broken. But the $180$ degree isospin rotations which simply interchange
$u$ and $d$ flavors remains intact. This symmetry breaking cannot result
from a $ud$ quark mass difference which would break both charge symmetry
as well as charge independence.  The alternative symmetry breaking by
an electromagnetic interaction is now preferred.

\subsection {Attempts to use the $ud$ mass difference to explain
isospin breaking}

If isospin is broken by the $ud$ mass difference as in the
``deuson" \cite{nils} model for the $X(3872)$ resonance,
one peculiar feature of the deuson wave function allows
the prediction (\ref{als}) for charged $B$ decays
even when isospin is broken. The $\{\bar{D^o}(\bar c
u)\,D^{*o}(c\bar u)\}$ wave function has no $d$ quarks,
nor $\bar d$ antiquarks. This enables its decay via the
meson state $V_u$ and suppresses its decay via the meson
state $V_d$. The state $V_u$ is an equal mixture of $\rho$
and $\omega$. But since the OZI-allowed final states for
$B^o$ decays (\ref{KM12}) have a $\bar d$
antiquark, we see that the neutral $B$ decays into the
$X(3872)$ resonance are forbidden in the deuson model.

If isospin is conserved, both the charged and neutral $B$ mesons should decay
equally into the $X(3872)$ resonance. Any isospin breaking will destroy this
equality. The deuson model completely suppresses production of $X$ in decay of neutral $B$
meson.
Thus the recent observation of neutral $B$ decays into the $X$ raises serious problems
for any of the proposed models for the $X(3872)$ resonance.

If both the $B^+$ and $B^o$ go into
both $J/\psi \rho$ and
$J/\psi \omega$ via the $X$ and the $X$ has no charged state,
there is a serious problem.

There are four relevant hadronizations of the final states
$c\bar s \bar c \qq$ state produced
in the dominant $\,b\,$ quark decay
$\bar b \rightarrow  \bar s c  \bar c$, followed by the creation of an
additional isoscalar $q \bar q $ pair from gluons.
\begin{itemize}
\item[(a)]
$ B^+(\bar b u)\,\rightarrow\,  c\bar s \bar c u \,\rightarrow\,
(\bar s u) \, J/\psi (\bar u u)$

\item[(b)]
$ B^+(\bar b u)\,\rightarrow\,  c\bar s \bar c u \,\rightarrow\,
(\bar s d) \, J/\psi (\bar d u)$

\item[(c)]
$ B^o\,(\bar b d)\,\rightarrow\,  c\bar s \bar c d \,\rightarrow\,
(\bar s u)\, J/\psi (\bar u d)$

\item[(d)]
$ B^o\,(\bar b d)\,\rightarrow\,  c\bar s \bar c d \,\rightarrow\,
(\bar s d)\, J/\psi (\bar d d)$
\end{itemize}
where the quarks in final state four-quark system have been grouped
to obey the OZI rule.
In the isospin symmetry limit, all four diagrams are equal.

Experiments show that diagrams (a) and (d) are both present and
approximately equal, while diagrams (b) and (c) produce charged $X$
states that are not seen. Diagrams (a) and (d) go into one another by the
$u\leftrightarrow d$ interchange which is an isospin reflection.

The problem is to find a symmetry breaking mechanism which breaks isospin
by suppressing (c) and (d), while keeping the isospin reflection that relates
(a) and (d).
The deuson model or any other model which suppresses the $d$ quark contribution
by the quark mass difference suppresses (d) while keeping (a).
The quark mass difference cannot suppress (b) and (c) while keeping (a) and (d).

We now note how these conclusions are changed if diagrams that violate the
OZI rule are included. Since these have the isoscalar $q \bar q$ appearing in
the same final state, they cannot produce charged states. The only
vector meson they can produce is the isoscalar $\omega$.
The amplitude for this transition can
interfere with the much larger OZI-allowed amplitude that produces the
$\omega$ but it cannot produce the $\rho$.
\section {The Solution; An isoscalar tetraquark with electromagnetic isospin breaking}

One possible direction\cite {ABW} for resolving this puzzle is to
consider the $X(3872)$ to be an isoscalar state with its strong decay
into $J/\psi \omega$ allowed, but suppressed by phase space because it
is so near the threshold. This explains why the $X(3872)$ is so narrow.
Experimental support for this conjecture comes from a recent BaBar analysis
of $B^+ \rightarrow J/\psi\, \omega$ which finds that the $\pi^+ \pi^- \pi^o$
invariant mass peaks more than $2\Gamma_\omega$
below the central value 783 MeV of $\omega$ mass
\cite{delAmoSanchez:2010jr}.
The $J/\psi \rho$ decay is isospin forbidden in strong interactions, but can
occur via an electromagnetic transition. This decay can compete with the
isospin allowed $J/\psi \omega$ decay because it has a
much higher phase space.
The $\rho$ is so wide that the $J/\psi \rho$ decay amplitude
goes far below the $J/\psi \omega$ threshold.

One possible model for the X(3872) is a $c\bar c q\bar q$ tetraquark
which is an isospin singlet. The analogous isotriplet $c\bar c q\bar
q$ tetraquark can decay strongly into $J/\psi \rho$ with such a large
width that it will never be seen.

Other models like the deuson model\cite{nils} note that the mass of the $X(3872)$
is close to the $D^{*+}D^-$ threshold and consider isospin breaking by the $ud$
mass difference at this threshold. That the $D^{*+}D^-$
and $J/\psi \omega$ thresholds are so close has no simple known
explanation. In the tetraquark model the $D^{*+}D^-$
threshold plays no role while the closeness of the $X(3872)$
to the $J/\psi \omega$ thresholds is crucial.

The tetraquark model for isosinglet $X(3872)$ also
explains the approximately
equal production from charged and neutral $B$ decays.
There is no simple
relation between the $J/\psi \omega$ and $J/\psi \rho$ decays. They are
roughly equal because one is allowed by strong interactions
but has little phase space, while
the other is forbidden and has much larger phase space. This picture
is not contradicted by the data.

All molecular models are in trouble
because no $D^*\bar D$ model can explain how isospin is broken in
a way that conserves the $u\leftrightarrow d$ interchange. The main
point here is the closeness of the $X$ to the $J/\psi \omega$ threshold,
not the $D^*\bar D$ threshold.

In this picture the only isospin breaking is in the electromagnetic
transition $X \rightarrow J/\psi \rho$. Thus all isospin relations for
$X$ production should
be valid and can be tested by experiment.
Since the decay $b\rightarrow  c \bar c s $ is a $\Delta I = 0$
transition
the final state of a  $B$ decay into charmonium and a kaon decay must have
the isospin 1/2 of the spectator nonstrange quark. This means that the  $B$
decays to $\,J/\psi K \rho\,$  are  related by isospin
Clebsch-Gordan coefficients to give the experimentally testable relation
\beq{Krhoiso}
\begin{array}{ccc}
& \phantom{AAAA} A[B^- \rightarrow J/\psi \bar K^o \, \rho^-]  & =
\\
= & {-}\sqrt {2} \cdot A[B^- \rightarrow J/\psi K^- \rho^o] & =
\\
= &{-}\sqrt {2} \cdot A[\bar B^o \rightarrow J/\psi \bar K^o\, \rho^o] & =
\\
= & \phantom{AAAA} A[\bar B^o \rightarrow J/\psi K^- \rho^+ ] &
\end{array}
\eeq

The mass for the $c\bar c u\bar u$ tetraquark has been estimated
\cite{ditri}
to be only 4\% larger
than the mass of two $D$
mesons in a wave function where the two quarks are coupled
to a color sextet and the two antiquarks are coupled to a color antisextet.
This suggests that such states with the color
sextet-antisextet coupling should
be included in all calculations for tetraquark states.
One crucial feature of the sextet coupling in this tetraquark model is that the interactions within the quark pair and the antiquark pair
are repulsive and are overcome by the four attractive quark-antiquark interactions. The mean distance between the two quarks or two antiquarks
is larger then the mean quark-antiquark distances\cite{triex,Nambu}. These color couplings and color-space correlations are not found in most other
tetraquark models.

The sextet-antisextet  state is found to have considerably a considerably lower mass\cite{ditri} than the
commonly used triplet-antitriplet diquark-antidiquark\cite{diqdiqbar,eric} models. This casts doubt on
all tetraquark calculations for the X(3872) resonance which neglect the color space
correlations\cite{diqdiqbar,eric,Hogaasen}. These neglect the basic physics seen in the
experimental hadron mass spectrum and its description by QCD motivated
models\cite{NewPenta,JW,DGG,Jaffe,Lipflasy} showing that the attractive
$q\bar
 q$ interaction as observed in mesons is much stronger than the
attractive $qq$
 interaction observed in baryons.

The color-space correlation contributions to the energy may well be more
important than  the color-magnetic energy neglected here which dominates other
tetraquark model calculations\cite{diqdiqbar,eric,Hogaasen}.

The $c u \bar c \bar u$ tetraquark which
contains charge-conjugate sextet-antisextet states have charge
conjugation quantum numbers that are conserved in strong and
electromagnetic decays. Starting with the spin quantum numbers, we define the
notation where the spins of the quark and the antiquark are respectively denoted
by
$s_q$ and $s_{\bar q}$ and the total spin is denoted by
$S$. The $\ket{s_q,s_{\bar q};S}$ states
 $\ket{1,1;0}$, $\ket{1,1;2}$ and
$\ket{0,0;0}$  are all even under $C$ and therefore also even
under $CP$. The $\ket{1,1;1}$ state is odd under $C$ and therefore
also odd under $CP$, while the $\ket{0,1;1}$ and $\ket{1,0;1}$ are
linear combinations of even and odd $C$ and the sum and difference
of these states are respectively even and odd under both $C$ and
$CP$.

In order to make contact with experiment, we recall that observed decay modes of $X(3872)$
are $J/\psi \rho^o$, $J/\psi \omega$, $J/\psi \gamma$ and $\bar D D^*$. The mode which
is conspicuously absent is $\bar D D$, which has a significantly larger phase space than
both $J/\psi \rho^o$ and $J/\psi \omega$. In order to understand this pattern, we need to
examine the constraints from the $C$ and $CP$ quantum numbers.

The $J/\psi \rho^o$, $J/\psi \omega$ and $J\psi\gamma$ are all even under $C$
and even under $CP$ (in $S$-wave).
The only
tetraquark state that is even under $C$ and cannot decay into $D \bar D$ is the
even $C$ linear combination of $\ket{0,1;1}$ and $\ket{1,0;1}$.
The $J^P = 1^+$, even-$C$ linear combination of $\ket{0,1;1}$ and $\ket{1,0;1}$
can decay into  $J/\psi \rho^o$ and $J/\psi \omega$  in an
$S$-wave,  and cannot decay into $D \bar D$, since a $1^+$ state cannot go into two
pseudoscalars. This $J^{PC}=1^{++}$ tetraquark model agrees with
all the published data for the $X(3872)$.

\section {Conclusion}

The apparent isospin violation in X(3872) decays is explained by an isoscalar tetraquark
model which conserves isospin, with the only isospin breaking arising from the electromagnetic
transition to $J/\psi \rho$. This decay is isospin forbidden for strong interactions,
but has much higher phase space than $J/\psi \omega$ because the $\rho$ is so wide and the
$J/\psi \rho$ decay amplitude goes far beyond the $J/\psi \omega$ threshold.
Models which break isospin by the $ud$ quark mass differences are unable to explain the data.
\section*{Acknowledgements}
This work was supported by the U.S. Department of Energy, Office of Nuclear
Physics, under contract DE-AC02-06CH11257.
We are grateful to
Barry Wicklund for pointing out the possibility of an isospin violating EM decay of
the $X(3872)$. We also thank
Arafat Mokhtar for interesting discussions of the data.
\vfill\eject
} 


%
\catcode`\@=11 
\def\references{
\ifpreprintsty \vskip 10ex
%
\hbox to\hsize{\hss \large \refname \hss }\else
\vskip 24pt \hrule width\hsize \relax \vskip 1.6cm \fi \list
{\@biblabel {\arabic {enumiv}}}
{\labelwidth \WidestRefLabelThusFar \labelsep 4pt \leftmargin \labelwidth
\advance \leftmargin \labelsep \ifdim \baselinestretch pt>1 pt
\parsep 4pt\relax \else \parsep 0pt\relax \fi \itemsep \parsep \usecounter
{enumiv}\let \p@enumiv \@empty \def \theenumiv {\arabic {enumiv}}}
\let \newblock \relax \sloppy
 \clubpenalty 4000\widowpenalty 4000 \sfcode `\.=1000\relax \ifpreprintsty
\else \small \fi}
\catcode`\@=12 
{\tighten

 } 
\end{document}